# High dynamic range electro-optic dual-comb interrogation of optomechanical sensors

D. A. Long,[1,*] B. J. Reschovsky,[1] T. W. LeBrun,[1] J. J. Gorman,[1] J. T. Hodges,[1] D. F. Plusquellic,[2] J. R. Stroud[2]

[1]*National Institute of Standards and Technology, 100 Bureau Dr, Gaithersburg, MD 20899, USA*
[2]*National Institute of Standards and Technology, 325 Broadway, Boulder, CO 80305, USA*
*\*Corresponding author: david.long@nist.gov*

**An interleaved, chirped electro-optic dual comb system is demonstrated for rapid, high dynamic range measurements of cavity optomechanical sensors. This approach allows for the cavity displacements to be interrogated at measurement times as fast as 10 μs over ranges far larger than can be achieved with alternative methods. While the performance of this novel readout approach is evaluated with an optomechanical accelerometer, this method is applicable to a wide range of applications including temperature, pressure, and humidity sensing as well as acoustics and molecular spectroscopy.**

Cavity optomechanical sensors offer exceptional sensitivity to a wide range of different physical parameters including force, pressure, acoustics, and humidity [1]. For example, a Fabry-Pérot microcavity has been used within an optomechanical accelerometer to achieve acceleration resolution below 1 μm/s² with a bandwidth greater than 13 kHz [2]. Similarly, a photonic crystal cavity has been integrated into a nanophotonic waveguide to measure temperature over a 100 K range [3]. These sensors require rapid and high dynamic range measurements of a cavity mode that is often coupled to the displacement of a mechanical resonator. However, widely utilized optical locking techniques generally limit these applications to measurement of small displacements and have never been demonstrated with a dynamic range approaching $10^5$ [4-6].

Recently, we demonstrated an alternative approach based upon the use of electro-optic frequency combs to record a given optical cavity mode during high-speed displacement (i.e., cavity length changes) [7]. This approach relied on a self-heterodyne [8, 9] architecture to down-convert a 2.4-GHz-wide optical frequency comb into the radiofrequency (RF) domain for digitization. Here, we have increased the comb bandwidth by an order of magnitude to 22.2 GHz by utilizing an interleaved dual comb (multiheterodyne) system based on two chirped waveforms to perform the optical down-conversion [10, 11]. The chirped waveforms provide a robust method for obtaining flat comb amplitudes [9-11] in contrast to other electro-optic comb methods that use overdriven, cascaded electro-optic modulators (EOM) [12] or injection locking schemes [13]. For high-speed displacement measurements, the use of repeated linear chirps to precisely define the bandwidth and comb resolution is key to optimizing the measured cavity response function while maximizing data throughput.

A schematic of the interleaved dual comb spectrometer is shown in Fig. 1. The stabilized external-cavity diode laser was split into sample and local oscillator (LO) legs having a 200 kHz relative frequency shift. The sample leg EOM was driven by a train of linear frequency chirps [9] from an arbitrary waveform generator (AWG) where the sample leg is swept from $f_{start}$ = 10 MHz to $f_{stop}$ = 11.1 GHz in the duration $\tau_{CP} = 1$ μs, and the LO is swept from $f_{LO_{start}} = 0$ Hz to $f_{LO_{stop}} = 10.09$ GHz in the same 1 μs duration. This resulted in a sample optical frequency comb having a span of 22.2 GHz and a comb tooth spacing of 1 MHz. This optical frequency comb is reflected off the optical cavity and then down converted into the RF domain by interfering with the LO optical frequency comb.

Fig. 2a shows how the differential chirps of the sample and LO interact to produce a comb in the RF region that carries the spectral information from the sample. The RF comb has a spacing of $1/\tau_{CP} = 1$ MHz and a span given by $f_{RF_{start}} = f_{start} - f_{LO_{start}} = 10$ MHz and $f_{RF_{stop}} = f_{stop} - f_{LO_{stop}} = 1010$ MHz for the majority of the results shown. The sample waveform is repeated $N_{chirps}$ times while the LO is phase shifted by a term scaled by $1/(\tau_{cp}N_{chirps})$ to separate the higher-order sidebands (harmonics) present in the down-converted RF comb [10]. The optical spectral ranges of 10 MHz to 11.1 GHz on both sides of the laser's center frequency are mapped down to the RF range from 10 MHz to 1.01 GHz. As illustrated in Fig. 2b, the positive and negative sidebands (+/-) are separated by twice the difference in the acousto-optic modulators' frequencies ($2\Delta f_{AOM}$) while the EOM orders are sequentially shifted by $1/(\tau_{cp}N_{chirps})$. With a chirp duration of 1 μs, the resulting 1st order RF comb has 1000 comb teeth per (+/-) sideband at a resolution of 1 MHz which corresponds to an effective optical resolution of 11.09 MHz.

The microwave power used to drive the EOM was selected to generate a flat comb spectrum while maximizing the power in the first order comb. The power dependence of the EOM response is compensated by decreasing the AWG microwave power for the lower frequency components. The residual EOM response causes some fluctuation in the power of the comb lines (see upper panel of Fig. 3), resulting in the variation of SNR seen in the lower panel of Fig. 3. The maximum dynamic range of the programmed waveforms is defined by the 8-bit AWG.

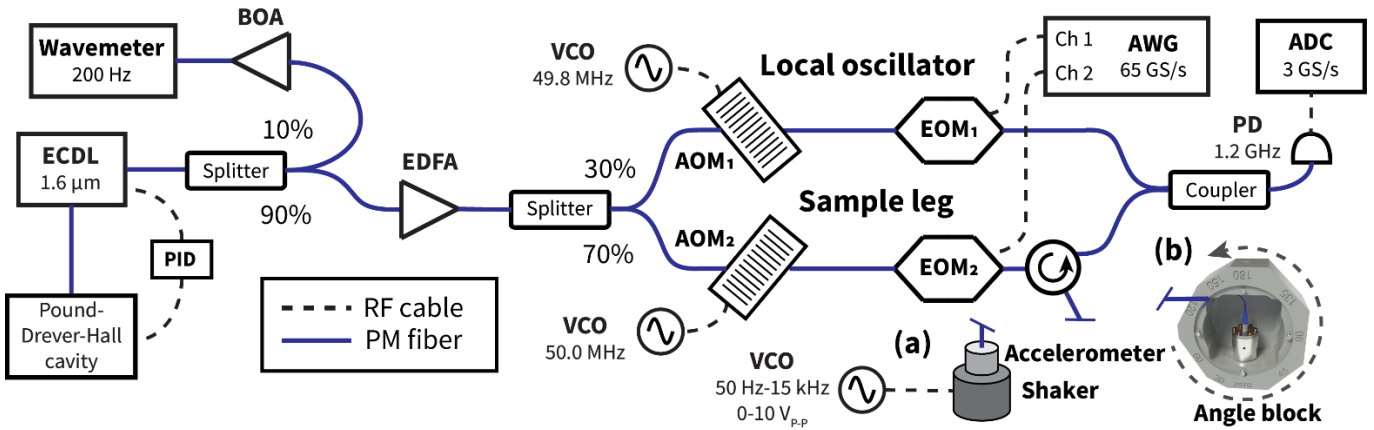

**Fig. 1.** Schematic diagram of the electro-optic dual comb spectrometer. The external cavity diode laser (ECDL) is stabilized to a reference cavity via a proportional-integral-derivative (PID) servo. The beam is split in fiber into sample and local oscillator paths where electro-optic frequency combs are generated via tailored chirped waveforms. The sample comb interrogates the optomechanical accelerometer in reflection before the beams are recombined on to a photodetector (PD). The accelerometer was placed on either an electromechanical shaker table (a) or within an aluminum angle block (b). Other abbreviations are: booster optical amplifier (BOA), erbium-doped fiber amplifier (EDFA), acousto-optic modulator (AOM), voltage-controlled oscillator (VCO), electro-optic phase modulator (EOM), arbitrary waveform generator (AWG), and analog-to-digital converter (ADC

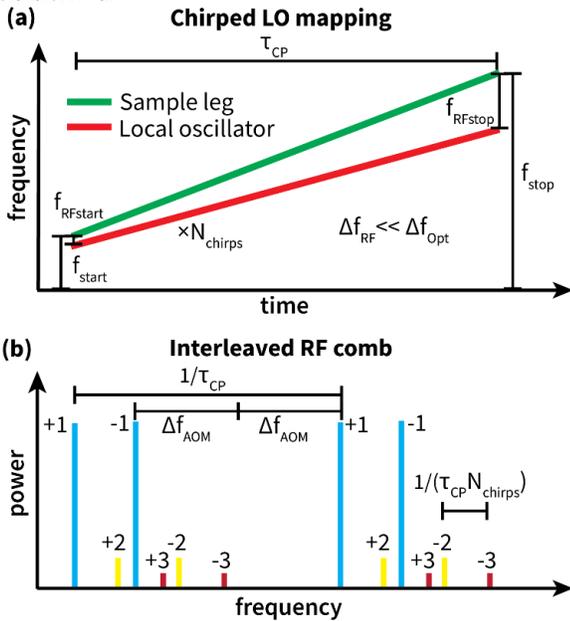

**Fig. 2.** (a) The spectrogram of the sample in green and local oscillator in red. The spectral information from the sample is mapped into the RF region defined by the difference in chirp rates. (b) An illustration of the interleaved (+/-) sidebands and sideband orders (harmonics) in the RF region (not to scale).

The device under test for these measurements was a fiber-coupled, cavity optomechanical accelerometer which has previously been characterized in detail [2, 7]. Briefly, the accelerometer is based upon a plano-concave Fabry-Pérot microcavity in which the flat mirror is suspended by a series of microscale beams. This design leads to an isolated mechanical resonance located at 7.849(3) kHz with a quality factor of 15.9(2) in air. The optical cavity was approximately 240 µm in length with a finesse near 2200 at 1600 nm.

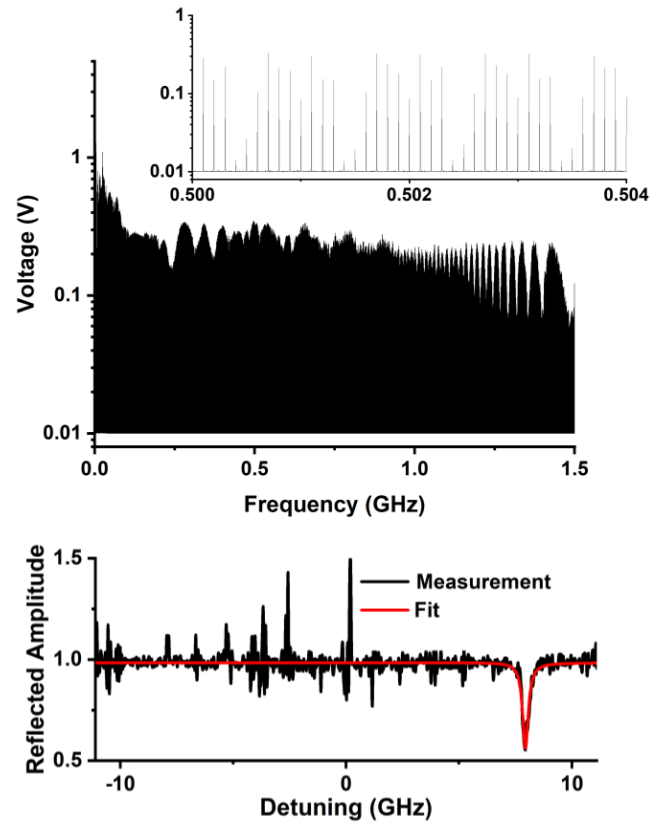

**Fig. 3.** (Upper panel) Typical dual comb RF spectrum where the inset shows a magnification. (Lower panel) Typical optical cavity spectrum and corresponding Fano resonance fit. Variations in comb SNR are due to the residual EOM frequency response and interference with unwanted reflections.

As shown in Fig. 1, the laser is amplified to 16 mW by an erbium doped fiber amplifier (EDFA) to maintain sufficient power of 2.5 mW at the sample without producing unwanted intermodulation products at the detector. The reflection from the microcavity and the LO leg are combined and then mixed on a 1.2 GHz photodetector (PD) that is digitized by a 3 GS/s analog to digital converter (ADC).

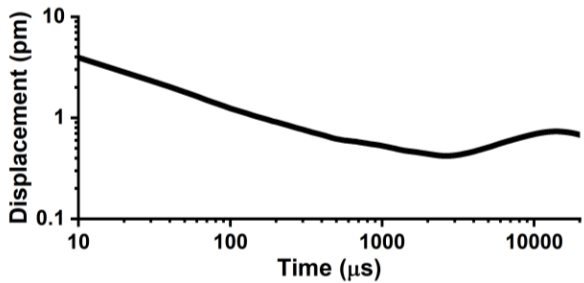

**Fig. 4.** Overlapped Allan deviation [14, 15] calculated from a noise-floor measurement taken when the optomechanical accelerometer was not excited.

For each measurement, a 0.1 s long interferogram was recorded (300 MS in length). Each of these interferograms was divided into 10-µs-long sections (i.e., 30 kS) and Fourier transformed to produce ten thousand 1st-order comb spectra (upper panel of Fig. 3). We note that the interleaving scheme separates the comb teeth of the EOM orders which provides a path to the even larger bandwidth coverage [10]. The magnitudes of the Fourier transform at each known comb tooth frequency were normalized against a comb spectrum recorded when the laser was detuned from the cavity resonance. The resulting cavity mode spectra were then fit with a Fano line shape [16] with the cavity mode center frequency as the only floated parameter (lower panel of Fig. 3). The typical center frequency fit uncertainty in the absence of any external excitation was 3 MHz, corresponding to a displacement sensitivity of 12 fm/Hz$^{1/2}$ (see Fig. 4 for a representative Allan deviation). When the accelerometer experiences an acceleration, the flat mirror of the microcavity translates, changing the cavity length. This displacement can subsequently be converted to acceleration using a damped harmonic oscillator model of the isolated mechanical resonance [2].

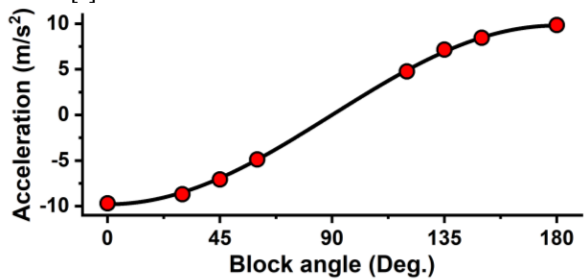

**Fig. 5.** Measured average acceleration as a function of block angle (red circles) for a series of repeated measurements. We note the standard deviation was smaller than the symbols. The measured displacement was referenced to the cavity mode position when the block was at 90°. Also shown is the calculated acceleration due to the known local gravity (black solid line) [17].

To assess the accuracy of the electro-optic dual comb readout method, the optomechanical accelerometer was mounted in a custom fabricated aluminum enclosure (see Fig. 1). This angle block allowed for the accelerometer to be rotated between known angles with respect to local gravity. As can be seen in Fig. 5 rotation of the accelerometer gives rise to an acceleration change of 2$g$ (1 $g \approx$ 9.81 m/s$^2$) as the block traverses 180°, corresponding to a cavity mode shift over 6 GHz. Based upon a fit of the measured acceleration, local gravity in Boulder, Colorado was measured to be 9.85(3) m/s$^2$, which is within 2σ of the known value [17]. We note that this accuracy is significantly better than that typically demonstrated with cavity optomechanical accelerometers [18, 19]. The achieved accuracy is likely limited by thermal drifts induced by manually rotating the block, which we expect could be significantly reduced with a mechanized apparatus.

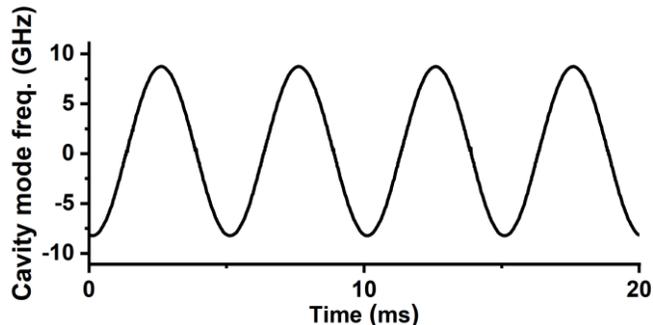

**Fig. 6.** Cavity mode frequency shift when the electromechanical shaker table was driven with a 200 Hz sine wave. The peak-to-peak frequency shift of 17 GHz corresponds to sixty cavity mode widths (full-width at half-maximum). Note that the y-axis range corresponds to the width of the optical frequency comb.

The optomechanical accelerometer was also mounted on an electromechanical shaker table which was driven by a sinusoidal voltage to assess the dynamic performance of the dual comb optical readout. As shown in Fig. 6, the comb readout readily measured a sinusoidal cavity mode frequency shift across a range of 17 GHz. This dynamic change in the microcavity's resonance frequency, which corresponds to sixty cavity mode widths, is far in excess of that generally possible with a frequency locking based approach or the previously demonstrated single-comb approach [7]. The high dynamic range of the dual comb readout method allowed for the linearity of the accelerometer and shaker table to be examined over a wide range of accelerations (see Fig. 7). The standard deviation of the residuals of a linear fit to this data was 0.06 m/s$^2$, where the maximum acceleration measured is 19 m/s$^2$.

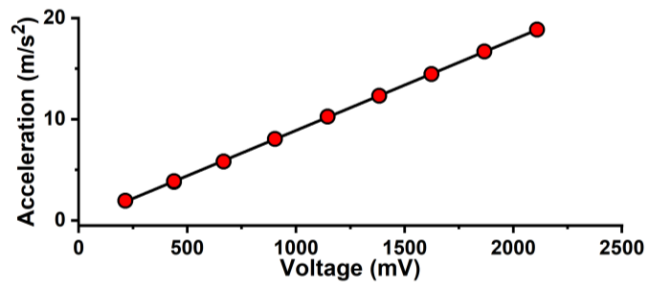

**Fig. 7.** Root-mean-squared acceleration as a function of electromechanical shaker drive voltage at 200 Hz. The standard deviation of the shown linear fit to this data was 0.06 m/s$^2$.

The frequency response of the optomechanical accelerometer was assessed by varying the frequency of the shaker excitation at constant drive voltage (see Fig. 8). The mechanical resonance of the optomechanical accelerometer can clearly be seen near 8 kHz as well as resonances in the shaker below 1 kHz and near 14 kHz.

Over the course of these measurements, we observed asymmetry in the line shape that depended on both the displacement of the resonance and the chirp direction (see Fig. 9). Rapid passage signals arise from interference between the induced transient field of a resonant oscillator, occurring at frequency, $f_r$, and a dynamically varying sample field exhibiting a distribution of frequencies $f_{ch}(t)$ in the vicinity of $f_r$. Examples include transient responses from a molecular system [10, 11], Fabry-Perot cavity [20], or atomic magnetic resonance [21] to chirped radiation.

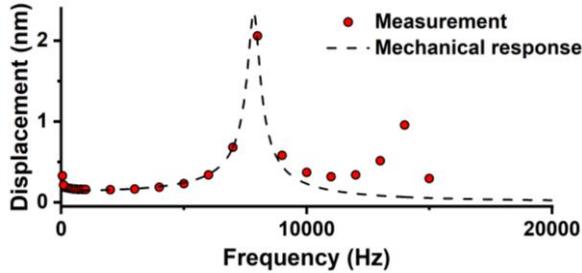

**Fig. 8.** Root-mean-squared displacement as a function of frequency for an electromechanical shaker drive of 1 V. Also shown is the fitted accelerometer mechanical resonance based upon a thermal noise measurement [22]. The second resonance found at 14 kHz is a mechanical resonance of the shaker table.

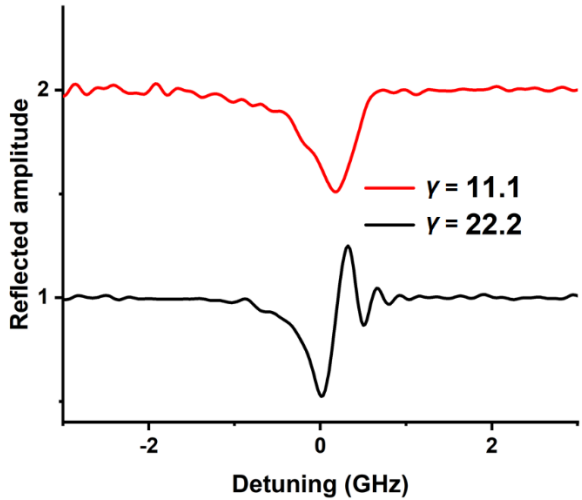

**Fig. 9.** Optical cavity modes measured with our typical sample chirp rate of 11.1 MHz/ns and two different LO chirp rates, resulting in the optical scan range to RF detection bandwidth ratios, γ, of 11.1 in red (offset for clarity) and 22.2 in black. As γ is increased, observe oscillations in the cavity mode indicative of rapid passage.

For chirped excitation of an optical resonator having a characteristic field storage time of $\tau$, rapid passage oscillations will become important when the frequency span of the sample laser chirp during the time interval $\Delta t = \tau$ exceeds the cavity halfwidth [20, 23]. This condition is met when the normalized frequency scan rate, $\nu_{ch} = 2\pi \dot{f}_{ch}\tau^2/n$, is greater than unity ($n$ is the index of refraction). Here the cavity storage time is $\tau = 1.136$ ns, giving a $\nu_{ch} = 0.09$, which is much below the expected threshold for rapid passage. However, the chirped LO comb magnifies the transient response at cavity resonance, $f_r$, by a factor $\gamma = \Delta f_{Opt}/\Delta f_{RF}$, which is the ratio of the optical scan bandwidth to the down converted RF bandwidth [10, 11] to give values of $\nu_{ch}$ near unity.

While we can avoid this rapid passage regime by limiting the chirp rate and maximizing detection bandwidth, we note that the dual comb readout's ability to probe these transient effects with high spectral resolution could have applications in the study of cavity dynamics and molecular systems [10, 11]. We also note that the described model provides only a semi-quantitative description, and that further analysis of these effects will be the subject of a future publication.

The electro-optic dual comb approach presented here is ideally suited to the rapid and high dynamic range interrogation of optical-cavity-based sensors as well as atomic and molecular species. Future work will focus upon further increasing the dynamic rage using higher-order sidebands, which can be readily separated in the RF domain through our interleaved dual-comb method. This should allow for cavity mode translations as large as 240 GHz to be recorded, corresponding to accelerations as large as $75g$ at 200 Hz.

**Acknowledgments.** This research was performed in part at the NIST Center for Nanoscale Science and Technology NanoFab. acknowledge G. E. Holland for his assistance in designing the angle block.

**Disclosures.** The authors declare no conflicts of interest.